\documentstyle[aps,pra,epsfig,twocolumn]{revtex}

\def\be{\begin{equation}}
\def\ee{\end{equation}}
\def\bea{\begin{eqnarray}}
\def\eea{\end{eqnarray}}
\def\bma{\begin{mathletters}}
\def\ema{\end{mathletters}}
\newcommand{\one}{\mbox{$1 \hspace{-1.0mm}  {\bf l}$}}
\def\C{\hbox{$\mit I$\kern-.7em$\mit C$}}

\begin{document}
\draft

\title{Entanglement capabilities of non--local Hamiltonians}

\author{W. D\"ur$^1$, G. Vidal$^1$, J. I. Cirac$^1$, N. Linden$^2$ and S. Popescu$^{3,4}$}

\address
{$^1$Institut f\"ur Theoretische Physik, Universit\"at Innsbruck,A-6020 Innsbruck, Austria\\
$^2$Department of Mathematics, University of Bristol, University Walk, Bristol BS8 1TW, UK\\
$^3$H.H. Wills Physics Laboratory, University of Bristol, Tyndall Avenue, Bristol BS8 1TL, UK\\
$^4$BRIMS, Hewlett-Packard Laboratories, Stoke Gifford, Bristol BS12 6QZ, UK}

\date{\today}

\maketitle

\begin{abstract}
We quantify the capability of creating entanglement for a general
physical interaction acting on two qubits. We give a procedure for
optimizing the generation of entanglement. We also show that a
Hamiltonian can create more entanglement if one uses auxiliary
systems.
\end{abstract}


\narrowtext

% --------------------------------------------------------------

In the last 2-3 years there has been very considerable increase in 
experimental activity aiming to create entangled quantum states.
One reason is the potential applications of entanglement to
quantum information processing.  Creating entanglement has been
possible in quantum optics for more than a decade, however now
many new communities, working in a variety of experimental areas
(for example NMR, condensed matter physics) are also joining the
field \cite{Fo00}. In general, entanglement between two systems
can be generated if they interact in a controlled way. However, in
most experiments these interactions are weak which make the
production of entanglement a very difficult task. Thus, it would
be very convenient to have a theory which would provide 
us with the best way of exploiting interactions to produce entanglement.

In this Letter we analyze the entanglement capabilities of Hamiltonians.
In particular, we would like to answer questions like: given an
interaction (Hamiltonian), what is the most efficient way of entangling
particles? Can we make the process more efficient by supplementing the
action of the Hamiltonian with some local unitary operations? Can we
increase the entanglement more efficiently by using some ancillas?

So far, much of the theoretical effort in quantum information
theory has been devoted to the characterization and quantification
of the entanglement of a given state.  Very recently, it has been
realised that there is a parallel notion of the entanglement in
the dynamics of a system \cite{Chefles}. In
\cite{Chefles}, the authors consider the situation that
one has a given unitary transformation and ask, for example, how
much state entanglement is needed to produce it. Here we focus on
a different issue: given an {\em interaction} (i.e. a Hamiltonian)
how can we make the most effective use of it\cite{Za00}. What we
propose here is to define and determine the entanglement
capabilities of physical processes, in particular, of unitary
evolutions \cite{Ci00}. This is a very relevant problem not only
from the theoretical point of view, but also from the experimental
one. Of course, this problem is even more difficult than the one
of quantifying the entanglement of states. In any case, in the
present work we give the first steps in this direction by
considering the case in which the physical process is acting on
two qubits.

From our results it turns out that: (i) It is more efficient to produce 
entanglement if one initially has already some; (ii) The best initial 
entanglement is universal, i.e. independent of the physical process; (iii) One 
can improve the performance of a physical process by complementing it with fast 
local operations; (iv) One can also improve it (in certain cases) by using 
auxiliary systems; (v) All entangling Hamiltonians can simulate each other and 
are thus qualitatively equivalent; we also
provide an upper bound on the time required for one Hamiltonian to
simulate another.

We consider two qubits interacting via a non--local Hamiltonian
$H$. We want to determine the most efficient way in which we can
use such an interaction to produce entanglement. We will
characterize the entanglement of a state of the qubits at a
given time $t$, $|\Psi(t)\rangle$, by some entanglement measure
$E$. In order to quantify the entanglement production, we define
the {\em entanglement rate} $\Gamma$ at a particular time $t$ of
the interaction as follows:
\be
\label{Gamma} \Gamma(t) \equiv \frac{dE(t)}{dt}. 
\ee 
This quantity depends on $|\Psi(t)\rangle$ not only through its entanglement 
$E$. The goal is then to find  the conditions which must be satisfied in order 
to obtain a maximal entanglement rate. In particular, we will be interested in 
determining the following:
\begin{description}
\item [(i)] 
 %{\it (i)} 
For any initial entanglement $E$ of the two-qubit system,
what is the state $|\Psi\rangle$, say $|\Psi_E\rangle$, for which
the interaction produces the maximal rate $\Gamma_E$.
\item [(ii)] 
 %{\it (ii)} 
The maximal achievable entanglement rate
$\Gamma_{\rm max}$ \cite{noteGamma},
$ %\be
\Gamma_{\rm max} \equiv \max_E \Gamma_E
$ %\ee
and the state $|\Psi_{\rm max}\rangle$ for
which $\Gamma=\Gamma_{\rm max}$.

\end{description}

These quantities are interesting because the knowledge of the
state $|\Psi_E\rangle$ will allow us to find out the most
efficient way of entangling the qubits. The idea is to supplement
the interaction Hamiltonian $H$ with appropriate local unitary
operations in such a way that the state of the qubits at any time
$t$ is precisely $|\Psi_{E(t)}\rangle$, for which the increase of
entanglement is optimal. In order to show how this can be
achieved, let us consider that the evolution given by $H$ proceeds
in very small time steps $\delta t$. Let us also assume that the
qubits are initially disentangled. Using local operations, we can
always prepare the state $|\Psi_0\rangle$
---that is, the product state which most efficiently becomes
entangled under the action of $H$. After a time step $\delta t$,
the state will change and its entanglement will increase to
$\delta E$. Then, we use (fast) local unitary operations to
transform the new state of the qubits into the state
$|\Psi_{\delta E}\rangle$ for which $\Gamma$ is optimal. Note that
this is always possible, since for qubits all states with the same
value of $E$, say $\delta E$, are connected by local unitary
transformations. By proceeding in the same way after every time
step, and taking the continuous time limit $\delta t\to 0$ we
obtain that the state of the qubits at time $t$ is always the
optimal one, $|\Psi_{E(t)}\rangle$. 
Obviously, in an experimental
realization, this procedure requires that we can apply the
appropriate local transformations in times which are short
compared to the typical time scale $\tau_H$ associated to $H$,
$ %\be
\tau_H =(e_{\rm max}-e_{\rm min})^{-1}, 
$ %\ee 
where $e_{\rm max}$ and $e_{\rm min}$ are the maximum and minimum eigenvalues of 
$H$, and we have set $\hbar=1$.

Knowledge of $\Gamma_{E}$ also permits us to determine the maximum
amount of entanglement $E_{\rm max}$ produced as a function of
time. We just have to express $\Gamma_E$ as an explicit function
of $E$, substitute it in (\ref{Gamma}) and solve that differential
equation to determine $E_{\rm max}(t)$. Note that the optimal
procedure described above will precisely reach the entanglement
$E_{\rm max}(t)$.

The state $|\Psi_{\rm max}\rangle$ is important since it gives
rise to the maximal increase of entanglement, and therefore
corresponds to the best operational point. After reaching the
state $|\Psi_{\rm max}\rangle$ with the procedure described in the
previous paragraph, the entanglement would be produced in a very
efficient way, if one could transfer the entanglement that is
gained after each time step $\delta t$ to other qubits (using
entanglement dilution \cite{Be96} or some other means). In particular, it
would increase proportionally to the time, $\Gamma_{\rm max}$
being the proportionality constant.

In the following, we will show how to determine $|\Psi_E\rangle$,
$\Gamma_E$, $|\Psi_{\rm max}\rangle$, and $\Gamma_{\rm max}$ for
an arbitrary Hamiltonian $H$. To this end, it is convenient to use
the Schmidt decomposition of the state of the qubits
$|\Psi(t)\rangle$ to write
\be
\label{Schmidt}
|\Psi\rangle = \sqrt{P} |\varphi,\chi\rangle +
e^{i\alpha} \sqrt{1-P} |\varphi^\perp,\chi^\perp\rangle,
\ee
where for the sake of short--hand notation we have omitted the time
dependence of all these quantities. Here,
$\langle \varphi|\varphi^\perp\rangle =
\langle \chi|\chi^\perp\rangle=0$ and $P\le 1/2$. 
 %In terms of the reduced density operators $\rho_{A,B}={\rm
 %Tr}_{B,A}(|\Psi\rangle\langle \Psi|)$, we have
 %\bma
 %\bea
 %\label{rhoA}
 %\rho_A |\varphi\rangle &=& P |\varphi\rangle,\\
 %\label{rhoB}
 %\rho_B |\chi\rangle &=& P |\chi\rangle.
 %\eea
 %\ema
Note that $E$ must only depend on the Schmidt coefficient $P$, given the fact
that it must be invariant under local unitary operations. For example, if we
choose as entanglement measure the entropy of entanglement \cite{Be96} --- the entropy of
the reduced density operator of one of the qubits---, we will have
\be
\label{vonNeumann}
E(P) = - P \log_2(P)- (1-P) \log_2 (1-P).
\ee
Note that the entropy of entanglement quantifies the amount of EPR entanglement
contained asymptotically in a pure state $|\Psi\rangle$. That is $E(P)$ gives
the ratio of maximally entangled EPR states
$|\Psi^-\rangle=1/\sqrt{2}(|01\rangle-|10\rangle)$ which can be distilled from
[are needed to create] $|\Psi\rangle$ respectively. Thus, we can write
\be
\label{Gamma2}
\Gamma(t) = \frac{dE}{dP} \frac{dP}{dt}.
\ee
In (\ref{Gamma2}), given a particular entanglement measure $E(P)$, we just
have to determine $dP/dt$. In order to do that, we need to find
the (infinitesimal) time evolution of the Schmidt coefficients
of the state of the qubits. After a time $\delta t$ we will have
$|\Psi(t+\delta t)\rangle=\exp(-iH\delta t)|\Psi(t)\rangle
\simeq (1- i H \delta t)|\Psi(t)\rangle$. The corresponding
reduced density operator $\rho_A(t+\delta t)$, where $\rho_{A,B}={\rm Tr}_{B,A}(|\Psi\rangle\langle \Psi|)$, can then be
written as
 %$\rho_A(t+\delta t) = \rho_A(t)
 %- i\delta t {\rm Tr}_B\{[H,|\Psi(t)\rangle\langle
 %\Psi(t)|]\}.$
$ %\be
\rho_A(t+\delta t) = \rho_A(t)
- i\delta t {\rm Tr}_B\{[H,|\Psi(t)\rangle\langle
\Psi(t)|]\}.
$ %\ee
The eigenvalues (Schmidt coefficients) of this operator can be
easily determined starting from %(\ref{rhoA})
$\rho_A |\varphi\rangle = P |\varphi\rangle$
and using standard perturbation theory. We find
\be
\label{dPdt}
\frac{dP}{dt} = 2 \sqrt{P(1-P)} \times {\rm Im}
[e^{i\alpha}\langle\varphi,\chi|H
|\varphi^\perp,\chi^\perp\rangle],
\ee
where we have omitted the time--dependence.
Upon substitution in (\ref{Gamma2}) we obtain the entanglement
rate. Since we are interested in maximizing $\Gamma$, it is
clear that we can always choose $\alpha$ such that
\be
\label{Gamma3}
\Gamma = f(P) |h(H,\varphi,\chi)|.
\ee
where
\bma
\bea
f(P) &=& 2 \sqrt{P(1-P)} E'(P),\\
 h(H,\varphi,\chi) &=&
\langle\varphi,\chi|H |\varphi^\perp,\chi^\perp\rangle.
\eea
\ema

By analyzing Eq.\ (\ref{Gamma3}) we can extract some interesting
conclusions, even before determining the maximum value of $\Gamma$
explicitly. Given the fact that $f$ and $h$ depend on different
parameters, in order to determine the quantities mentioned in
(i--ii) we can maximize the functions $f$ and $|h|$ independently.
First, if we want to determine the quantities mentioned in (i),
we have to fix the value of $E$. In that case, $P$ is also fixed
and therefore the maximum of the entanglement rate will correspond
to a state of the form (\ref{Schmidt}) with some fixed
$|\varphi\rangle$, $|\chi\rangle$, and $\alpha$ (which maximize
$|h|$). That is, for any value $E$ of the entanglement, the states
$|\varphi\rangle$ and $|\chi\rangle$ for which the maximal
entanglement rate $\Gamma_E$ is obtained do not depend on $E$, but
only on the form of the Hamiltonian $H$. Let us denote by $h_{\rm
max}$ the maximum value of $|h|$; that is,
\be
\label{hm} h_{\rm max} = \max_{||\varphi||,||\chi||=1}
|\langle\varphi,\chi|H |\varphi^\perp,\chi^\perp\rangle|. 
\ee
Then, we can easily determine how the entanglement would evolve
with time if we always drive the qubits with local operations so
that at each time their state corresponds to the optimal one. We
can simply solve the differential equation (\ref{dPdt}), obtaining
$P(t)= \sin^2[h_{\rm max}t+\phi_0],$ 
with $P(0)=\sin^2(\phi_0)$.
Using the explicit dependence of $E$ on $P$, we can directly then
calculate $E(t)$. The evolution of the entanglement is fully
characterized by $h_{\rm max}$, which is a quantity that only
depends on the interaction Hamiltonian. That is, for a given $H$,
$h_{\rm max}$ measures the capability of creating entanglement. In
the following we give a simple way of determining $h_{\rm max}$,
which allows us to classify the {\em entanglement capability} of
any  Hamiltonian. On the other hand, once the entanglement measure
is specified, we can calculate the value $P_0$ of $P$ for which we
obtain the maximal rate by simply considering the function $f(P)$.
For example, choosing the expression (\ref{vonNeumann}) for the
entanglement, we find that $P_0$ solves the equation
$ %\be
\ln \frac{1-P_0}{P_0}=\frac{2}{1-2P_0}, 
$ %\ee 
i.e. $P_0\simeq
0.0832$ which gives $E(P_0)\simeq 0.413$. This shows that, in
order to increase the entanglement of a two-qubit system in an
optimal way,  it is better to start with some initially entangled
state rather than a product state \cite{noteTurin}. Note that the optimal initial
entanglement $E(P_0)$ is independent of $H$. %the Hamiltonian.

In the following we will show how to determine the entanglement
capability $h_{\rm max}$ of a general Hamiltonian $H$ acting on
the qubits. First, we will show how, if we allow to supplement the
evolution of $H$ by local unitary operations, we can express $H$
in a standard form that only depends on three parameters. Then we
will derive an expression for $h_{\rm max}$ in terms of those
parameters.

Except for a trivial constant, we can always express a general
Hamiltonian as
\begin{eqnarray}
\label{Hgen} H &=&  \sum_{i=1}^3 \alpha_i \sigma_i^A\otimes
\one_B+
 \sum_{j=1}^3 \beta_j \one_A \otimes \sigma_j^B+
 \nonumber\\
 & &\qquad\sum_{i,j=1}^3 \gamma_{i,j} \sigma^A_i
\otimes \sigma^B_j.
\end{eqnarray}
Here, $\sigma_i$ are the Pauli operators, and $\vec\alpha$,
$\vec \beta$, and $\gamma$ are two real vectors and a real
matrix, respectively. 
We now show that by supplementing the evolution operator with
local unitary operations we can obtain an effective Hamiltonian
which has one of the two standard forms
 %\be
 %\label{Htilde} \hat H = \sum_{k=1}^3 \mu_k \sigma^A_k \otimes
 %\sigma^B_k,
 %\ee
\be \label{Htilde} \hat H^{\pm} = \mu_1 \sigma^A_1 \otimes
\sigma^B_1 \pm \mu_2 \sigma^A_2 \otimes \sigma^B_2 +\mu_3
\sigma^A_3 \otimes \sigma^B_3, \ee where  $\mu_1\ge \mu_2 \geq
\mu_3 \geq 0$ are the (sorted) singular values of the matrix
$\gamma$ \cite{foot8}. We first note that the terms corresponding
to $\vec \alpha, \vec \beta$ in (\ref{Hgen}) give no contribution
to $h_{\rm max}$ (\ref{hm}) and can therefore be neglected.
Second, we apply the local operations $U$ ($V$) and $U^\dagger$
($V^\dagger$) to the first (second) qubit at the beginning and end
of the evolution process, respectively. We select them such that
$ %\be
U^\dagger\sigma_i^A U = \sum_{k=1}^3 O^A_{k,i} \sigma^A_k, %\quad
V^\dagger\sigma_j^B V = \sum_{l=1}^3 O^B_{j,l} \sigma^B_l,
$ %\ee
where $O^{A,B}$ are orthogonal matrices of determinant one, each
being plus or minus the orthogonal matrices in a singular value
decomposition of $\gamma$. Thus the total (non-local) effect of
the evolution for a time $t$ is equivalent to the one obtained
with the Hamiltonian $\hat H^+$ ($\hat H^-$) for the same time if
det$\gamma \geq 0$ (det$\gamma <0$). Without loss of generality,
we may take $H$ of the form $\hat H^+$ (\ref{Htilde}) in what
follows \cite{noteH+}.

Now let us determine $h_{\rm max}$ in terms of $\mu_{1,2,3}$. We
can write
$ %\be
h(H,\varphi,\chi) = \sum_{k=1}^3 \mu_k
\langle\varphi|\sigma^A_k|\varphi^\perp\rangle \langle
\chi|\sigma^B_k|\chi^\perp\rangle. 
$ %\ee 
Using the Cauchy--Schwarz
inequality, it can be  checked that the maximum of (the absolute
value of) this function is reached for
$|\chi\rangle=|\varphi^\perp\rangle$. In this case, using the fact
that $|\varphi\rangle\langle \varphi|+
|\varphi^\perp\rangle\langle \varphi^\perp|=\one$ we obtain
$ %\be
h(H,\varphi,\varphi) = \sum_{k=1}^3 \mu_k - \sum_{k=1}^3 \mu_k
\langle\varphi|\sigma_k|\varphi\rangle^2. 
$ %\ee 
Taking into account
that  $\mu_1\ge \mu_2\ge \mu_3$, we see that the maximum value
occurs when $|\varphi\rangle=|0\rangle$ or
$|\varphi\rangle=|1\rangle$, i.e. an eigenstate of $\sigma_3$. For
that choice we obtain
\be
h_{\rm max}=\mu_1+\mu_2. 
\ee
Summarizing, once we have transformed the Hamiltonian 
$H$ to the
standard form (\ref{Htilde}) we obtain that for a given value of
$E$ (and therefore of $P$), 
\bma \label{total} 
\bea 
|\Psi_E\rangle &=& \sqrt{P} |0,1\rangle + i \sqrt{1-P} |1,0\rangle,\\ \Gamma_E
&=& f(P) h_{\rm max} 
\eea 
\ema 
where $h_{\rm max}=\mu_1+\mu_2$. The maximum rate $\Gamma_{\max}$ is obtained 
for $P=P_0$, where $P_0$ is the value that maximizes $f(P)$. Thus, $|\Psi_{\rm 
max}\rangle$ and $\Gamma_{\rm max}$ are given by (\ref{total}) with $P=P_0$. For 
example, for the entanglement measure (\ref{vonNeumann}), $P_0\simeq 0.0832$ 
which leads to $f(P_0) \simeq 1.9123$.

So far, we have calculated the most efficient way of entangling
two qubits if we can use local unitary operations acting on each
of the qubits. We have not allowed, however, local operations
which entangle each of the qubits with local ancillas. We will
now show that this possibility permits us to increase the
maximum entanglement rate $\Gamma_{\rm max}$ for certain kind of
Hamiltonians. We will first generalize the formulas derived
above to the case of multilevel systems, given that the system
qubit--plus--ancilla is of this sort. We consider a state
$|\Psi\rangle$ with Schmidt decomposition
 $%\be
|\Psi\rangle = \sum_{n=1}^N \sqrt{\lambda_n} |\varphi_n,\chi_n\rangle.
$ %\ee
As before, any entanglement measure $E$ will only depend on the
Schmidt coefficients $\lambda_n\ge 0$. In particular, in the
following we will use the entropy of entanglement,
$ %\be
E(\vec \lambda) = -\sum_{n=1}^{N} \lambda_n \log_2 (\lambda_n).
$ %\ee
Using the definition (\ref{Gamma}) of entanglement rate, we have
\be
\label{Gamma4} \tilde\Gamma = \sum_{n=1}^N \frac{\partial
E}{\partial \lambda_n} \frac{d\lambda_n}{dt} = \frac{1}{N}
\sum_{n,m=1}^N \left[ \frac{\partial E}{\partial \lambda_n} -
\frac{\partial E}{\partial \lambda_m} \right]
\frac{d\lambda_n}{dt}, \ee
where we have used the fact that the
sum of all the Schmidt coefficients is constant. Using
perturbation theory as before, we find 
 %\bea \frac{d\lambda_n}{dt}
 %&=& 2 \sqrt{\lambda_n} {\rm Im} [\langle
 %\varphi_n,\chi_n|H|\Psi\rangle]\\ &=& 2 \sum_{m=1}^N
 %\sqrt{\lambda_n\lambda_m} {\rm Im} [\langle
 %\varphi_n,\chi_n|H|\varphi_m,\chi_m\rangle].\nonumber 
 %\eea
$\frac{d\lambda_n}{dt}
= 2 \sqrt{\lambda_n} {\rm Im} [\langle
\varphi_n,\chi_n|H|\Psi\rangle] = 2 \sum_{m=1}^N
\sqrt{\lambda_n\lambda_m} {\rm Im} [\langle
\varphi_n,\chi_n|H|\varphi_m,\chi_m\rangle].$

Rather than proceeding in complete generality we now consider an
example which demonstrates that adding ancillas may allow one to
increase entanglement more efficiently than is possible without
the use of ancillas.  We will consider the case in which the
ancillas are also qubits. We write $P=\lambda_1$ and concentrate
on the case in which $\lambda_2=\lambda_3=\lambda_4=(1-P)/3$. In
that case, Eq.\ (\ref{Gamma4}) simplifies to
\be
\label{Gamma5} \tilde\Gamma = \tilde f(P) \tilde
h(H,\varphi_n,\chi_n) \ee where now \bma \bea & &\tilde f(P) = 2
\sqrt{P(1-P)/3} \log_2[(1-P)/(3P)],\label{tf}\\ & & \tilde
h(H,\varphi_n,\chi_n) = \sum_{n=2}^4 {\rm Im}
[\langle\varphi_1,\chi_1|H|\varphi_n,\chi_n\rangle]. \eea \ema We
can always choose the phase of the states $|\varphi_n\rangle$ such
that all the terms on the sum add with the same sign. We can
therefore replace the imaginary parts of the terms in the above
expression by their absolute values, and in (\ref{Gamma5}) we can
replace $\tilde f(P)$ by $|\tilde f(P)|$. We find that $\tilde P_0
\simeq 0.8515$ (which corresponds to an entropy of entanglement
$E(\tilde P_0) \simeq 0.8415$) maximizes $|\tilde f(P)|$
(\ref{tf}) and leads to $|\tilde f(\tilde P_0)| \simeq 1.6853$.
Proceeding as before, we can easily maximize $\tilde h$. We obtain that
the maximum value is
$ %\be
\tilde h_{\rm max} = \mu_1 + \mu_2 + \mu_3,\label{thm} 
$ %\ee 
which occurs when $|\varphi_n\rangle=|\chi_n\rangle$ are orthogonal maximally 
entangled states between the qubit and the ancilla. For example in the case that $\det \gamma\geq 0$, the choice
$|\varphi_1\rangle=|\phi^+\rangle,
|\varphi_2\rangle=i^{\frac{3}{2}}|\psi^+\rangle,
|\varphi_3\rangle=i^{\frac{1}{2}}|\psi^-\rangle,
|\varphi_4\rangle=i^{\frac{3}{2}}|\phi^-\rangle,$
where
$\{|\phi^\pm\rangle,|\psi^\pm\rangle\}$ are Bell states
\cite{noteBell}, together with $P=P_0=0.8515$ leads to a maximal
[under the previous assumptions on the $\lambda_i$'s] entanglement
rate $\tilde\Gamma = \tilde\Gamma_{\max}$.

Let us compare the cases in which we use ancillas and the one in
which we do not use them. On the one hand, we have that $|\tilde
f(\tilde P_0)| < |f(P_0)|$. But on the other, $\tilde h_{\rm
max}\ge h_{\rm max}$. Thus, if $\mu_3\ne 0$ it may be the case
that the use of ancillas can help to increase the maximum rate of
entanglement $\Gamma_{\rm max}$ as well as the rate $\Gamma_E$ for
a given entanglement $E$ of state $|\Psi\rangle$. This is in fact
the case, if we have, for example, $\mu_1=\mu_2=\mu_3$ (i.e.
$\tilde h_{\rm max}=3/2 h_{\rm max}$). In this case we obtain
$\tilde \Gamma_{\max} \simeq 1.3220 \Gamma_{\max}$. In a similar
way, one can check for this specific Hamiltonian that $\tilde
\Gamma_E \geq \Gamma_E$ if the initial entanglement satisfies $E
\geq 0.08$.

 %--- 
Finally, it is easy to 
show that all entangling Hamiltonians are {\it 
qualitatively} equivalent when assisted by local operations. In particular, 
given two Hamiltonians $H$ and $H'$ with either $h_{\rm max}=\alpha h'_{\rm max}$ or 
$\tilde h_{\rm max}=\alpha \tilde h'_{\rm max}$, one can simulate the action of $H'$ 
for any time $t$ by applying $H$ for at most $3\alpha^{-1}t$. This can be seen as follows: 
Applying $H$ of the form (\ref{Htilde}) for $\delta t/2$ followed by a local 
unitary operation $\sigma_1$ in $A$ before and after another application of $H$ for $\delta 
t/2$ is equivalent to the application of the Hamiltonian $H_1=\mu_1 
\sigma_1\otimes\sigma_1$ for the time $\delta t$, provided that $\delta t$ is 
infinitessimally small. Since $H_1$ is locally equivalent to 
$H_k=\mu_1\sigma_k\otimes\sigma_k$, applying sequentially $H_k$ for $\delta t 
\mu'_k/\mu_1$ is equivalent to the application of $H'$ for the time $\delta t$. 
Using the restrictions on $\mu_k,\mu'_k$ the claim readily follows. The %important 
question of efficient simulation of another 
Hamiltonian in the same time $t$ will be addressed in %some 
future work. %\cite{Vi01}. 
%---

In summary, we have found the optimal way of using any non--local
interaction to entangle a pair of qubits. The idea is to use local
operators to drive the instantaneous state to the one that
maximizes the entanglement rate, at each moment of the evolution.
We have found that the  entanglement %-producing 
capacity of any given Hamiltonian is determined by the sum of the two largest
singular values of the matrix $\gamma$ defined in (\ref{Hgen}).
Finally, we have shown that for certain Hamiltonians one can
overcome this maximal entanglement rate by using ancillas prepared
in maximally entangled states with the qubits.

We are grateful to C. H. Bennett for very valuable input, particularly
concerning the issue of the qualitative equivalence of Hamiltonians.
This work was supported by the Austrian SF, the European Community (TMR
network ERB--FMRX--CT96--0087 ; project EQUIP (contract IST-1999-11053)), the
ESF, and the Institute for Quantum Information GmbH. G.V also acknowledges
funding from EC through grant HPMF-CT-1999-00200.

{\it Note added:} The idea of universal simulation of Hamiltonians as discussed 
in this manuscript originated in a larger collaboration, including several other 
people and is developed in detail in \cite{Be01}. 
After completion of this manuscript we have also learned that the notion of 
universal simulation of Hamiltonians has been independetly addressed by Dodd et al. \cite{Do01}.

% -------------------------------------------------------------


\begin{references}
\bibitem{noteGamma}
Note that the definitions of $\Gamma$ and $\Gamma_{\rm max}$ are not restricted to qubits but are also valid for $d$--level systems, $d\geq2$.
\bibitem{foot8} Given a real matrix $\gamma$ one
can always find two orthogonal matrices $O^1$ and $O^2$ such that
$\gamma=O^1 \mu O^2$, where $\mu$ is a diagonal matrix with
positive entries; this is called a singular value decomposition of
$\gamma$ and the diagonal elements of $\mu$ are the singular
values.
\bibitem{noteH+}
Due to the fact that $\hat H^-=\hat H^{+^{T_A}}$, where $T_A$ denotes the partial 
transposition with respect to system $A$, we have that $\langle\varphi,\chi|\hat H^+ 
|\varphi^\perp,\chi^\perp\rangle = \langle\varphi^{\perp *},\chi|\hat H^- 
|\varphi^{*},\chi^\perp\rangle$. That is, if $h_{\rm max}$ for $\hat H^+$ is obtained 
by states $|\varphi,\chi\rangle$, $h_{\rm max}$ for $\hat H^-$ is obtained by the 
states $|\varphi^{\perp *},\chi\rangle$. Thus, we can restrict our analysis to $\hat H^+$.
\bibitem{noteBell}
The Bell states are defined as
$|\phi^\pm\rangle=1/\sqrt{2}(|00\rangle\pm|11\rangle)$ and
$|\psi^\pm\rangle=1/\sqrt{2}(|01\rangle\pm|10\rangle)$.
\end{references}
\end{document}